\documentclass[aps,prl,showpacs,twocolumn]{revtex4}
\usepackage{mathrsfs}
\usepackage{amssymb}
\usepackage{amsmath}
\usepackage{bm}
\usepackage{graphics}
\usepackage{natbib}

\begin{document}
\title{SPT 2-Channel Kondo Model in the Structure of Normal Metal/Quantum Dot/\emph{DIII}-class Topological Superconductor}

\author{Zhen Gao}
\email[Current address: Department of Physics, University of Konstanz, Germany. Email address: ] {tomgaophysik@gmail.com}
\author{Wei-Jiang Gong}
\email[Email address: ] {gwj@mail.neu.edu.cn}

\affiliation{College of Sciences, Northeastern University, Shenyang 110819, China}
\date{\today}

\begin{abstract}
We investigate the Kondo effect in a structure which is constructed by embedding one quantum dot between a normal metal and a \emph{DIII}-class topological superconductor supporting Majorana doublets at its ends. It is observed that Kondo correlation occurs between the localized state in the dot and two continuum states simultaneously, i.e., the continuum state in the metal and the continuum Andreev reflection state between the metal and topological superconductor. As a result, the Kondo model Hamiltonian is topologically protected by the $SU(2)_s\rtimes Z_2^T$ symmetry. More interestingly, two new Kondo temperatures appear in this system, in comparison with the normal Kondo model. This phenomenon exactly reflects the special role of Majorana doublet in tuning the Kondo effect.
\end{abstract}

\maketitle

\emph{Introduction}. ---Topological superconductor (TS), a kind of fermionic symmetric protected topological(SPT) state system, has attracted extensive interest in the field of mesoscopic physics because Majorana modes appear at the ends of the one-dimensional TS which are of potential application in topological quantum computation.\cite{Majorana1,RMP1,Fuliang1} Owning to the presence of Majorana zero mode, its various transport behaviors have been observed, such as fractional Josephson effect, the resonant Andreev reflection, and so on.\cite{Majorana2,Alice} Especially in the \emph{DIII}-class TS, which is time-reversal-invariant, Majorana mode appears in pairs due to Kramers's theorem.\cite{Zhang2,Fuliang2,Ryu,Qi,Teo,Timm,Beenakker2} As a result, at each end of the \emph{DIII}-class TS nanowire there will exist one Majorana doublet.\cite{Deng,Nakosai,Wong,Zhang,Nagaosa2,Sunk} It causes the period of Josephson current to be dependent on the fermion parity of the \emph{DIII}-class TS junction.\cite{Liuxj} Also interestingly, Josephson effect presents apparent time-reversal anomaly phenomenon in the junction formed by the coupling between the \emph{DIII}-class TS and $s$-wave superconductor.\cite{Qixl}
\par
The other notable characteristics of Majorana mode are its influence on the transport behaviors of the non-topological systems, especially the quantum-dot (QD) systems, via coupling t the electronic bound state. When Majorana mode connects laterally with one closed QD circuit, the conductance magnitude will be suppressed by one half.\cite{LiuDE} If it couples to two QD sub-circuits serially, nonlocal phenomenon can be achieved with its potential application in nonlocal quantum entanglement.\cite{DoubleQD}
Alternative influence of Majorana mode on the QD systems consists in its nontrivial role in modifying the electron correlation mechanism, especially the Kondo physics. Due to the fundamental importance of this topic, it has received much attention.\cite{cite2,cite1,cite3} For a typical structure where Majorana zero mode couples to the metal via one KQD, the Kondo physics induces various new results. It has been observed that in addition
to the Kondo fixed point, the system flows to a new fixed point controlled by the Majorana-induced
coupling, which is characterized by the correlations between KQD and the fermion parity of the TS and metal.\cite{cite3} Besides, the interplay between Kondo effect and Andreev reflection modifies the conductance to a great extent.\cite{cite1}
\par
In view of the published works, Majorana mode indeed plays an important role in changing the Kondo effect. However, all of them focus on the influence of Majorana mode in $D$-class TS. It is necessary to investigate the effect of Majorana mode in the $DIII$-class TS for presenting a complete analysis about this issue. Note, also, that because of its time-reversal symmetry, the Majorana mode here (i.e., Majorana doublet) will modify the Kondo effect from a new aspect. With this idea, in the present Letter we aim to investigate the influence of Majorana doublet on the Kondo effect, by considering one structure formed via the indirect coupling between a normal metal and Majorana doublet via a KQD. Our calculation results show that in this structure, two different Kondo physics mechanisms coexist which originate from the antiferromagnetic correlations between the KQD and metal as well as that between the KQD and continuum Andreev reflection state respectively. As a consequence, three Kinds of Kondo temperatures come into being respectively in the presence of different couplings between the KQD and Majorana doublet. This result embodies the essential influence of Majorana doublet on the Kondo physics.

\begin{figure}[htb]
\begin{center}\scalebox{0.55}{\includegraphics{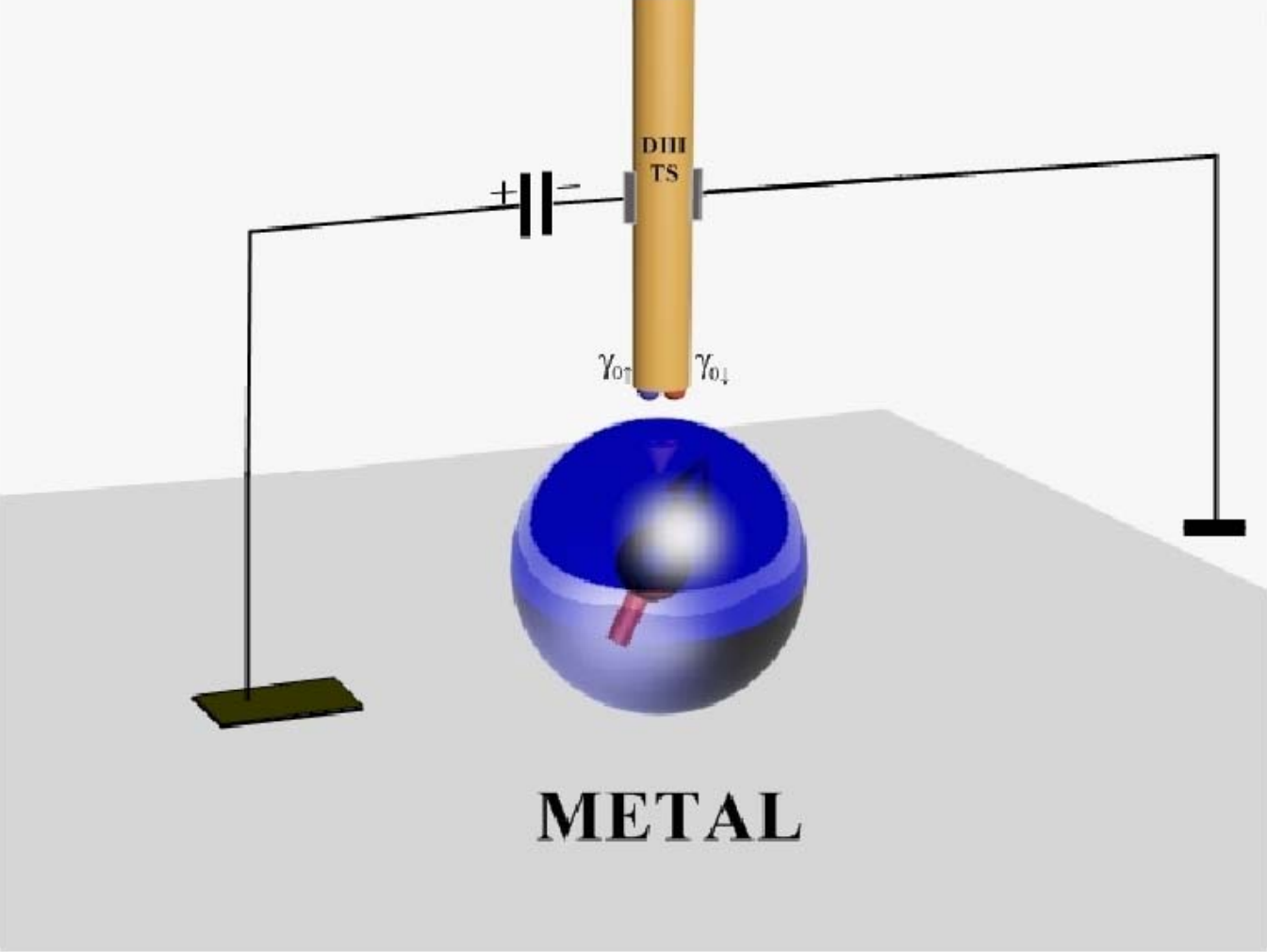}}
\caption{(a) Schematic of a structure of normal metal/KQD/\emph{DIII}-class TS.
\label{Struct}}
\end{center}
\end{figure}

\emph{Model}.
---The considered structure is illustrated in
Fig.\ref{Struct}, where a one-dimensional \emph{DIII}-class TS couples to one normal metal via one KQD. The system Hamiltonian can be
written as $H=H_0+H_D+H_{p}+H_I$. The first two terms denote the
Hamiltonians of the normal metal and the KQD, which take the forms
$H_0=\sum_{k\sigma}\varepsilon_k c^\dag_{k\sigma}c_{k\sigma}$ and $H_D=\varepsilon_d\sum_\sigma d^\dag_\sigma d_\sigma+U n_\uparrow
n_\downarrow$.
$c^\dag_{k\sigma}$ and $d^\dag_{\sigma}$ ($c_{k\sigma}$ and $d_{\sigma}$) are the electron creation (annihilation) operators in
the normal metal and KQD, respectively, with spin index $\sigma$. $\varepsilon_{k}$ is the electron energy at state $|k\sigma\rangle$ in
the normal metal. $\varepsilon_d$ is the energy level of the KQD, and $U$ denotes the intradot
Coulomb repulsion with $n_\sigma=d^\dag_\sigma d_\sigma$.
$H_p$ is the Hamiltonian of the $DIII$-class TS, which can be given by $H_p=-\mu_p\sum_{j\sigma}f_{j\sigma}^\dag f_{j\sigma}-
t\sum_{j\sigma}(f_{j+1,\sigma}^\dag
f_{j\sigma}+\text{h.c.})+\sum_{j\sigma\sigma'}
[(-i\sigma_1\sigma_2)_{\sigma\sigma'}\Delta_p f_{j+1,\sigma}^\dag f_{j\sigma}^\dag+\text{h.c.}]$\cite{Qixl,Qixl2}.
$f^\dag_{j\sigma}$ ($f_{j\sigma}$) is the electron creation (annihilation) operator in
the TS. $\mu_p$ and $t$
represent the chemical potential and intersite coupling in the \emph{DIII}-class TS. $\sigma_l$ ($l=1,2,3$) is the Pauli matrix. $\Delta_p$ is the
Copper-pair hopping term in the $DIII$-class TS.
Finally, $H_I$ describes the couplings between
the KQD and normal metal (TS). Its expression can be written as
$H_I=\delta
t\sum_\sigma f_{1\sigma}^\dag d_\sigma+\sum_{k\sigma}
V_k c_{k\sigma}^\dag d_\sigma +\text{h.c.}$,
where $\delta t$ and $V_k$ are the coupling
coefficients, respectively.
For the case of one infinitely long \emph{DIII}-class TS, one Majorana doublet
will form at its end, we can therefore project $H_p$ onto the
zero-energy subspace of $H_p$. As a result, $H$ is simplified as
\begin{small}
\begin{eqnarray}
H&=&\sum_{k\sigma}\varepsilon_k c_{k\sigma}^\dag
c_{k\sigma}+\varepsilon_d\sum_\sigma d_\sigma^\dagger d_\sigma+U n_\uparrow
n_{\downarrow}\notag\\
&&+(\sum_{k\sigma}V_k c_{k\sigma}^\dag d_\sigma+\sum_\sigma\delta t \gamma_{0\sigma}
d_\sigma+\text{h.c.}).\label{eq1}
\end{eqnarray}
\end{small}
$\gamma_{0\sigma}$ is the Majorana operator, which obeys Clidfford algebra
$\{\gamma_{0\sigma},\gamma_{0\sigma'}\}=2\delta_{\sigma\sigma'}$.
\par
For an Anderson-typed system, the Kondo physics can be well described by transforming it into one $s$-$d$ exchange model. Accordingly, by performing the Schrieffer-wolff transformation with anti-unitary operator $\mathcal{S}=S_0-S_0^\dagger$, where $S_0=\sum_{\sigma}[\sum_{k}V_k(\frac{1-n_{\bar{\sigma}}}{\varepsilon_k-\varepsilon_d}+\frac{ n_{\bar\sigma}}{\varepsilon_k-\varepsilon_d-U})c_{k\sigma}^\dagger d_\sigma+\delta t(\frac{1-n_{\bar\sigma}}{-\varepsilon_d} +\frac{n_{\bar\sigma}}{-\varepsilon_d-U})\gamma_{0\sigma}d_\sigma]$, we can write out the $s$-$d$ spin-exchange form of the Hamiltonian in Eq.(\ref{eq1}). Namely,
$H_{\rm eff}\approx H_0+\frac{1}{2}[S_0,H_I]=H_0+H_K+H_T$, in which
\begin{small}
\begin{eqnarray}
H_K=\sum_{j=1}^3 H^{(j)}_K&=&\sum_{kk',\sigma\sigma'}{\cal J}_{kk'}\boldsymbol{S}\cdot\boldsymbol{\sigma}_{\sigma\sigma'}c_{k\sigma}^\dagger c_{k'\sigma'}\notag\\
&&+\sum_{k,\sigma\sigma'} {\cal K}_k\boldsymbol{S}\cdot\boldsymbol{\sigma}_{\sigma\sigma'}(c_{k\sigma}^\dagger \gamma_{0\sigma'}+\gamma_{0\sigma}c_{k\sigma'})\notag\\
&&+{\cal M}\sum_{\sigma\sigma'}\boldsymbol{S}\cdot\boldsymbol{\sigma}_{\sigma\sigma'}\gamma_{0\sigma}\gamma_{0\sigma'}, \label{kondo1}
\end{eqnarray}
\end{small}
\par
with ${\cal J}_{kk'}=V_kV_{k'}(\frac{1}{\varepsilon_k-\varepsilon_d}+\frac{1}{\varepsilon_{k'}-\varepsilon_d}+\frac{1}{\varepsilon_d+U-\varepsilon_k} +\frac{1}{\varepsilon_d+U-\varepsilon_{k'}})$,
${\cal K}_k=V_k\delta t(\frac{1}{\varepsilon_k-\varepsilon_d}+\frac{1}{\varepsilon_d+U-\varepsilon_k}-\frac{1}{\varepsilon_d}+\frac{1}{\varepsilon_d+U})$, and
${\cal M}=\delta t^2(\frac{1}{\varepsilon_d+U}-\frac{1}{\varepsilon_d})$. $\textbf{S}$ is the spin momentum operator and $\boldsymbol{\sigma}$ is the pauli matrix. Eq.(\ref{kondo1}) shows that in addition to the normal Kondo term ($H^{(1)}_K$), two new spin-exchange interaction modes form due to the presence of Majorana doublet. However, the difference between them is well-defined. $H^{(2)}_K$ can be regarded as the antiferromagnetic correlation between the localized state in the KQD and the continuum Andreev reflection state which is formed between the metal and Majorana doublet. In contrast, the correlation in $H^{(3)}_K$ is virtual because Majorana doublet cannot confine one spin due to $\gamma_{0\sigma}^2=1$.
Next, $H_T$ describes the direct tunneling between the normal metal and Majorana doublet, which is given by
$H_T=\sum_{k,\sigma}{\cal W}_{k}c_{k\sigma}^\dagger\gamma_{0\sigma}+\text{h.c.}$
with ${\cal W}_{k}=V_k\delta t (\frac{1}{\varepsilon_k-\varepsilon_d}+\frac{1}{\varepsilon_k-\varepsilon_d-U}-\frac{1}{\varepsilon_d}-\frac{1}{\varepsilon_d+U})$. Such a term exactly reflects the screening effect of the Andreev reflection on the Kondo effect. Since the electron state near the Fermi level makes leading contribution to the spin correlation, $\varepsilon_k$ can be approximated to be zero in the parameters, i.e., ${\cal W}_k=-2V\delta t(\frac{1}{\varepsilon_d+U}+\frac{1}{\varepsilon_d})$. As a result, ${\cal W}_k$ will be suppressed near the position of electron-hole symmetry where the Kondo correlation is strengthened.

\emph{Structural symmetry}.
---The Hamiltonian in Eq.(\ref{kondo1}) exactly describes a 2-channel Kondo model with the spin-channel(flavor) mixing term. To understand its  properties, we need to discuss the symmetry of it first. For this purpose, we introduce a unified fermion field operator which labels both electrons and Majorana fermions with an extra channel index $j$, i.e.,
$\psi_{j\sigma}(x)=\left\{
  \begin{array}{cc}
    c_{\sigma}(x) & (j=1) \\
    \gamma_{0\sigma}(x) & (j=2)\\
  \end{array}
\right.$. Here $\gamma_{0\sigma}(x)=u_0(x)f_\sigma(x)+u^*_0f_\sigma^\dagger(x),\,|u_0|\sim e^{-|x|/\xi}$ ($\xi\ll 1$ because Majorana bound state decays exponentially fast into the bulk of superconductor).
Thereby, we can express the 2-channel topological Kondo model in $\boldsymbol{SU(2)}$ form:
$H=\int dx[v_j\psi^\dag_{j\sigma}i\partial_x\psi_{j\sigma}+J_K\delta(x)(\kappa^*\psi_{1\sigma}^\dag+\lambda^*\psi_{2\sigma}^\dag) \frac{\boldsymbol{\sigma}_{\sigma\sigma'}}{2}
(\kappa\psi_{1\sigma'}+\lambda\psi_{2\sigma'})\cdot\boldsymbol{S}]$.
In this equation, the energy scale is considered near the Fermi surface of metal, where the dispersion is linearized with two speeds, i.e., the Fermi speed $v_1=v_F$ and Majorana speed $v_2=v_M$. They are determined relatively by the chemical potentials of the metal and TS.
Surely, in the special case where $V_k=\delta t$, the Kondo Hamiltonian will get an isotropic form:
$J_K\delta(x)\tilde{\boldsymbol{s}}\cdot\boldsymbol{S}$
in which $\tilde{\boldsymbol{s}}=\frac{1}{2}(c_{\sigma}^\dagger+\gamma_{0\sigma}^\dagger)\boldsymbol{\sigma}_{\sigma\sigma'}(c_{\sigma'}+\gamma_{0\sigma'})$.
\par
It is known that for the normal 2-channel Kondo model, its Hamiltonian has the Sugawara form with the current operators of charge $U(1)_c$, spin $SU(2)_s$ and flavor(channel) $SU(2)_f$. The total Hamiltonian thus obeys the global gauge symmetry of $SU_2(2)_s\times SU_2(2)_f\times U(1)_c$ Kac-Moody algebra with total CFT central charge $c_{_{TOT}}=4$.\cite{cite4}
In our case, however, the spin-flavor mixing term appears with  $\psi_{j\sigma}^\dag(0)\tau^1_{jl}\boldsymbol{\sigma}_{\sigma\sigma'}\psi_{l\sigma'}(0)\cdot\boldsymbol{S}$, hence the global gauge symmetry is obviously broken. To clarify the Hamiltonian symmetry, we introduce a new spin-flavor superposition field operator to combine two channels into one: $\psi_{sf,\sigma}=\kappa\psi_{1\sigma}+\lambda\psi_{2\sigma}$ which possesses the anti-commutation rule $\{\psi_{sf,\sigma}(x),\psi^\dagger_{sf,\sigma'}(x')\}=\delta_{\sigma\sigma'}\delta(x-x')$ in the presence of proper choice of $\kappa$ and $\lambda$. It should be noticed that Majorana fermion has only half of the quantum degree of freedom compared with normal complex fermion with central charge $\frac{1}{2}$ while for complex fermion $c=1$. Thus under certain manipulation condition of $\frac{v_F}{v_M}=\frac{|\kappa|^2}{|\lambda|^2}$, the total Hamiltonian can also be rewritten into the Sugawara form: $H\sim\frac{1}{8\pi}J^2_{sf,c}+\frac{1}{6\pi}\boldsymbol{J}_{sf,s}^2+J_K\boldsymbol{J}_{sf,s}(0)\cdot\boldsymbol{S}$ with $J_{sf,c}=\overset{\cdot}{{}_{}{}_{\cdot}}\psi_{sf,\sigma}^\dag\psi_{sf,\sigma}\overset{\cdot}{{}_{}{}_{\cdot}}$ and $\boldsymbol{J}_{sf,s}=\overset{\cdot}{{}_{}{}_{\cdot}}\psi_{sf,\sigma}^\dag\frac{\boldsymbol{\sigma}_{\sigma\sigma'}}{2}\psi_{sf,\sigma'}\overset{\cdot}{{}_{}{}_{\cdot}}$.
At the stable Kondo fixed point, by absorbing impurity spin, the total spin current operator becomes $\tilde{\boldsymbol{J}}_{sf}(x)=\boldsymbol{J}_{sf,s}(x)+2\pi\delta(x)\boldsymbol{S}$ which obeys the commutation relation $[\tilde{J}_{sf}^a(x),\tilde{J}_{sf}^b(x')]=2\pi i\epsilon^{abc}\tilde{J}_{sf}^c(x)\delta(x-x')+\frac{3}{2}\delta^{ab}\pi i\frac{d}{dx}\delta(x-x')$, i.e., the $SU_{\frac{3}{2}}(2)_s$ Kac-Moody algebra.
This exactly means that our Hamiltonian behaves like the one-channel Kondo effect. It drives dynamically the fixed point of 2-channel Kondo model to where $c_{_{TOT}}=3$, because the flavor(channel) $SU(2)_f$ gauge symmetry breaks down in the presence of the spin-flavor mixing part.
\par
In addition, for our structure, both charge conservation $U(1)_c$ symmetry and even fermion parity symmetry $Z_2^p$ break down in TS due to the appearance of Majorana doublet. The reason is as follows. Since Majorana doublet obeys the time-reversal invariance, the emerging Majorana-Kondo coupling $\gamma_{0\sigma}\boldsymbol{\sigma}_{\sigma\sigma'}\gamma_{0\sigma'}\cdot\boldsymbol{S}$ is essentially a time-reversal anomaly term with $(i\gamma_{0\uparrow}\gamma_{0\downarrow})S^y\sim(-1)^{N_F}S_y$ depending on fermion number $N_F$ which labels the two-fold degenerate ground states by time-reversal-symmetric sector $Z_2$ index.\cite{Qixl} Therefore, our 2-channel Kondo model indeed possesses the $SU(2)_s\rtimes Z_2^T$ Kac-Moody gauge symmetry and therefore its ground states are robust to perturbations that do not violate the structural symmetry.
Such a conclusion can be verified by acting the Hamiltonian on the ground state which is formally expressed as $\sum_k g_k(c^\dag_{k\uparrow}|\downarrow\rangle-c^\dag_{k\downarrow}|\uparrow\rangle)|F\,\rangle\otimes(a|0\rangle_M+b|1\rangle_M)$. $|0\rangle_M$ and $|1\rangle_M$ are two basis vectors in Hilbert space generated by Majorana doublet. When calculating the ground state energy, one can find the virtual ``\emph{QD spin---conduction spin}" triplet pairing phenomenon induced by the proximity effect. It originates from the contribution of the superconduction pairing potential. However, in this system, the $p$-wave triplet pairing potential that leaks into QD is very weak near the Fermi level. As a result, the Kondo singlet pairing dominates with little effect of the virtual triplet pairing term on the change of the ground state energy.

\emph{Scaling and Kondo temperature}.
---In view of the new kind of Kondo correlation in $H^{(2)}_K$, we would like to discuss its contribution to the Kondo temperature of the whole system by means of the perturbation method. To do so, we define the the Green function that describes the scattering processes, i.e.,
$G_{c,\eta\eta'}(k\tau,k'\tau')=\sum_nG^{(n)}_{c,\eta\eta'}(k\tau,k'\tau')=-\sum_{n=0}^\infty\frac{(-1)^n}{n!}\int_0^\beta \cdots\int_0^\beta d\tau_1\cdots d\tau_n\langle T_\tau[H_K^{(2)}(\tau_1)\,\cdots$ $H_K^{(2)}(\tau_n)c_{k\eta}(\tau)c_{k'\eta'}^\dagger(\tau')]\rangle_0$. It is easy to find that the first-order scattering process is zero because the Wick contraction contains $\langle c^\dagger_{ks}\gamma_{0s'}\rangle_0$ and $\langle\gamma_{0s}c_{ks'}\rangle_0$. Therefore, $G^{(2)}_{c,\eta\eta'}(k\tau,k'\tau')$ makes the leading contribution. With the help of Wick theorem, $G^{(2)}_{c,\eta\eta'}(k\tau,k'\tau')$ can be expressed as the multiplication among free propagators, which is able to be described by the connected Feynman diagrams, as shown in Fig.2(a). In this figure, the solid, dashed, and dotted lines correspond to the free propagators $G^{(0)}_{c,\eta\eta'}(k,\omega_n)$, $G^{(0)}_{d,ss'}(\omega_m)$ and $G^{(0)}_{M,\sigma\sigma'}(\omega_l)$, respectively. The Feynman diagrams clearly show that the Andreev reflection achieves the Kondo effect. Meanwhile, in comparison with these two diagrams, the first scattering process can be found to contribute dominantly to the Kondo effect, since virtual processes occur in the second one.
So, it is adequate to calculate the contribution of the first Feynman diagram in Fig.2(a), to understand this kind of Kondo effect.
\par
According to the Dyson series in random-phase-approximation theory, the total particle-particle interactional vertex can be given by
\begin{small}
\begin{eqnarray}
&&\Gamma_{s_1\nu_1s_1'\nu_1'}(k_1,\omega_n;0,\omega_l;\omega_m)={{\cal K}\over2}(\boldsymbol{\sigma}_{s_1s_1'}\cdot\boldsymbol{\sigma}_{\nu_1\nu_1'})
+\notag\\
&&\frac{1}{\beta^2}\sum_{p,\omega_s,\omega_i}\sum_{s_2,\nu_2;s_2',\nu_2'}{{\cal K}\over2}(\boldsymbol{\sigma}_{s_2's_1'}\cdot\boldsymbol{\sigma}_{\nu_2'\nu_1'}) G^{(0)}_{c,s_2'}(p,\omega_s)\notag\\
&&G^{(0)}_{d,\nu_2'}(\omega_m-\omega_s)\times{{\cal K}\over2}(\boldsymbol{\sigma}_{s_2s_2'}\cdot\boldsymbol{\sigma}_{\nu_2\nu_2'})G^{(0)}_{M,s_2}(\omega_i)\notag\\
&&G^{(0)}_{d,\nu_2}(\omega_m-\omega_i)\times\Gamma_{s_1\nu_1s_2\nu_2}(k_1,\omega_s;0,\omega_i;\omega_m).\label{DysonG}
\end{eqnarray}
\end{small}
It can be clearly described by the Feynman-diagram illustration shown in Fig.2(b).
\begin{figure}[htb]
\begin{center}\scalebox{0.4}{\includegraphics{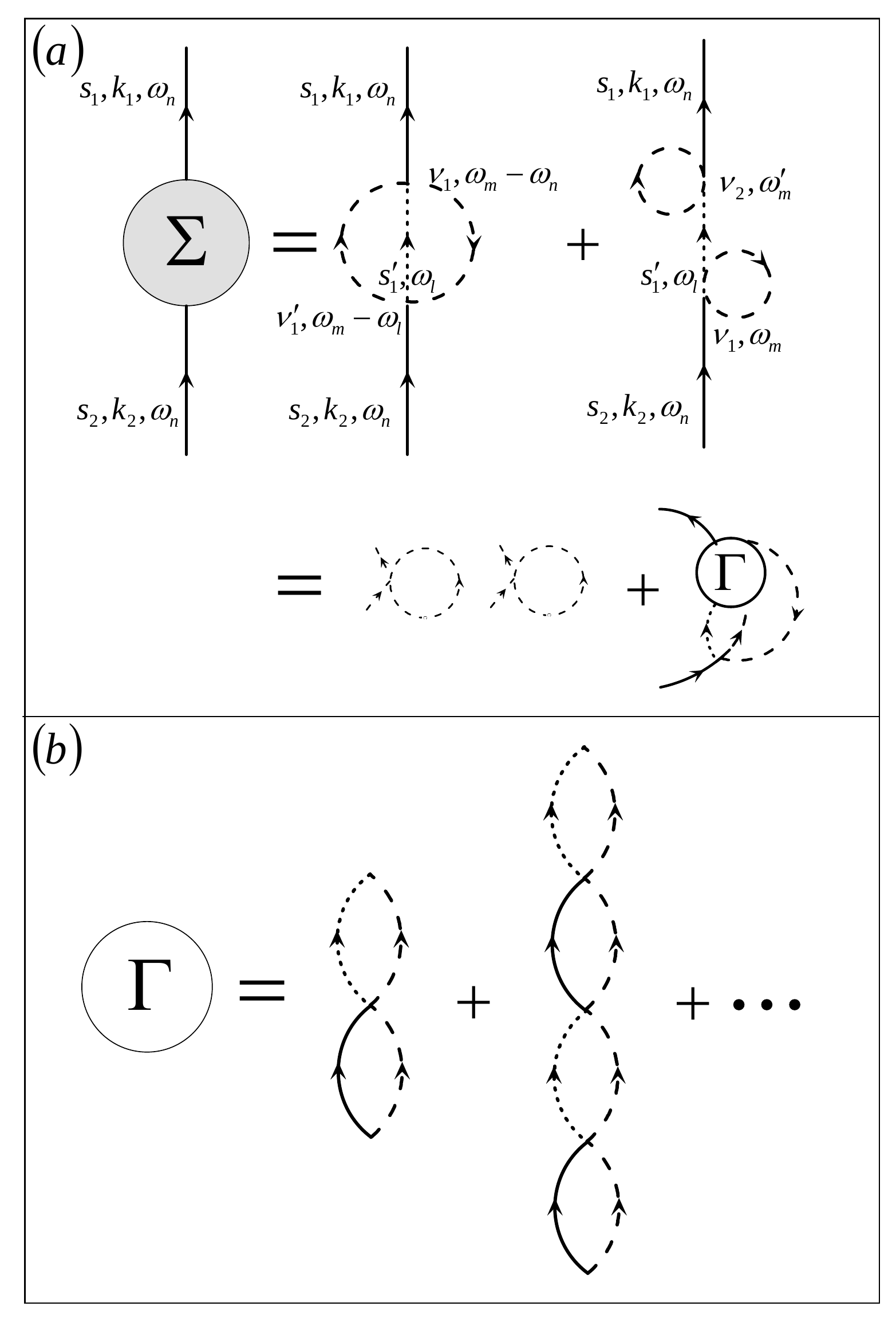}}
\caption{(a) Feynman diagrams of the second-order scattering process. (b) Illustration of the Dyson series of the vertex in the random-phase approximation.
\label{Feynman}}
\end{center}
\end{figure}
Next, by solving this equation, the property of $\Gamma$ will be clarified, as well as the leading property of the Kondo effect. Suppose
$\Gamma_{s_1\nu_1s_1'\nu_1'}=\Gamma^0\delta_{s_1s_1'}\delta_{\nu_1\nu_1'} +\Gamma^1(\boldsymbol{\sigma}_{s_1s_1'}\cdot\boldsymbol{\sigma}_{\nu_1\nu_1'})$ and apply the spin summation identity, we can figure out Dyson equation into two self-consistent equations of $\Gamma^0$ and $\Gamma^1$ in which two Kondo polarization bubbles
$\pi_1=\frac{1}{\beta}\sum_{p,\omega_s}G^{(0)}_c(p,\omega_s)G^{(0)}_d(\omega_m-\omega_s)$ and $\pi_2=\frac{1}{\beta}\sum_{\omega_i}G^{(0)}_M(\omega_i)G^{(0)}_d(\omega_m-\omega_i)$ are involved. And we employ the Fedatov-Popov method with introducing an imaginary chemical potential $\mu_f=-i\pi\frac{T}{2}$ ($T$ is temperature with $k_B=1$) to shift the Matsubara frequency into $\tilde{\omega}_m=\omega_m-\pi\frac{T}{2}=2\pi T(m-\frac{1}{4})$ in order to eliminate the nonphysical states in pseudofermion representation. Then precise to the lowest-order diagram in the vertex (for $H_K^{(2)}$, the two-loops), the scaling RG equation for $\cal K$ can be written out: $\beta^2(\mathcal{K})+\frac{5\rho\mathcal{K}^3}{|\varepsilon_d|}\beta(\mathcal{K})+\frac{9\rho^2\mathcal{K}^6}{4\varepsilon_d^2}=0$ with the beta function $\beta(\mathcal{K})=\frac{d\mathcal{K}}{d\ln D}$, where $D$ is the metal bandwidth acting as a UV cut-off.
Therefore, the Kondo temperatures contributed by $H^{(2)}_K$ can be defined, i.e.,
$
T^{(1)}_K\sim D e^{-\frac{|\varepsilon_d|}{9\rho{\cal K}^2}}=D e^{-\frac{1}{2\rho{\cal J}}\cdot {2|\varepsilon_d|\over 9 {\cal M}}}\,,\,
T^{(2)}_K\sim D e^{-\frac{|\varepsilon_d|}{\rho{\cal K}^2}}=D e^{-\frac{1}{2\rho{\cal J}}\cdot {2|\varepsilon_d|\over {\cal M}}}
$
under the condition of $\varepsilon_d<0$. As is known, the first term in Eq.(\ref{kondo1}) is the typical Kondo coupling whose one-loop RG equation has been well studied before: $\beta(\mathcal{J})=\frac{d\mathcal{J}}{d\ln D}=-2\rho\mathcal{J}^2$, which achieves the Kondo temperature $T^{(0)}_K\sim D e^{-\frac{1}{2\rho{\cal J}}}$. Thus, three kinds of Kondo temperatures appear respectively in this structure, with their explicit relationships. This also indicates that due to the coupling between KQD and Majorana doublet, the KQD level becomes an important quantity to influence the Kondo physics, in addition to $\cal M$.
\par
We next take the limit of electron-hole symmetry to present the competition of the Kondo temperatures. At the position of $\varepsilon_d\rightarrow-U/2$, ${2|\varepsilon_d|\over {\cal M}}\rightarrow({U\over 2\delta t})^2$, so the leading Kondo physics is determined by the value of $U/\delta t$. If $U>6\delta t$, the Kondo temperatures obey the relationship that $T^{(0)}_K>T^{(1)}_K>T^{(2)}_K$. This means that in the case of the weak coupling between the KQD and Majorana doublet, the antiferromagntic correlation between the KQD and metal still governs the Kondo physics. With the increase of $\delta t$ to the region $6\delta t>U>2\delta t$, the relation among the Kondo temperatures will be changed to be $T^{(1)}_K>T^{(0)}_K>T^{(2)}_K$. This result can be explained as follows. The increase of $\delta t$ enhances the Andreev reflection, so the correlation between the KQD state and Andreev reflection state becomes the leading mechanism. Next when $\delta t$ further increases to make the KQD and Majorana doublet form a molecule, the Kondo-temperature relationship will be $T^{(1)}_K>T^{(2)}_K>T^{(0)}_K$. Although such an analysis deviates from the physics picture within the perturbation framework, it should be understood that adjusting $\delta t$ can change the ground state of this structure and lead to the variation of Kondo mechanism. Up to now, the underlying physics of the Kondo effect has been clarified.

\par
In what follows, we evaluate the conductance feature induced by $H^{(2)}_K$. First, it is necessary to mention the particle-hole scattering process where $\omega_s\rightarrow-\omega_s$ and $\varepsilon_k\rightarrow-\varepsilon_k$. Due to the particle-hole symmetry of conduction electron density of states, we can get the result that $\pi_{1(2)}(\omega^{p-p})=\pi_{1(2)}(\omega^{p-h})$, as well as particle-hole vertex $\Upsilon^0(\omega^{p-h})=-\Gamma^0(\omega^{p-p})$ and
$\Upsilon^1(\omega^{p-h})=-\Gamma^1(\omega^{p-p})$. The total vertex can therefore be given by
$\Gamma_T(\omega)=\Gamma_{s_1\nu_1s_1'\nu_1'}(\omega^{p-p})+\Upsilon_{s_1\nu_1s_1'\nu_1'}(\omega^{p-h})
-{{\cal K}\over2}(\boldsymbol{\sigma}_{s_1s_1'}\cdot\boldsymbol{\sigma}_{\nu_1\nu_1'})$.
With the help of spin summation $\sum_{s_1'\nu_1'\nu_1}[\Gamma^0_T\delta_{s_1s_1'}\delta_{\nu_1\nu_1'} +\Gamma^1_T(\boldsymbol{\sigma}_{s_1s_1'}\cdot\boldsymbol{\sigma}_{\nu_1\nu_1'})](\boldsymbol{\sigma}_{s_1's_2}\cdot\boldsymbol{\sigma}_{\nu_1'\nu_1})
={3}\Gamma^1_T\delta_{s_1s_2}$, the conductance quasiparticle irreducible
self-energy can be expressed as
$
\Sigma(\omega_n)=\frac{3\cal K}{2\beta}\sum_{m\in\mathbb{Z}}\frac{\pi_{2}(\tilde{\omega}_m)\Gamma^1_T(\tilde{\omega}_m)}{i\omega_n-i\tilde{\omega}_m}.
$
After the frequency summation and evaluating $-2\text{Im}\Sigma_F$, we can arrive at the result of relaxation time which reflects tunneling conductance caused by the Kondo term $H^{(2)}_K$, i.e.,
$\tau_1^{-1}+\tau_2^{-1}\sim\frac{1}{\rho}(1+\ln\frac{T_K^{(1)}}{T})^{-2}+\frac{3}{\rho}(1+\ln\frac{T_K^{(2)}}{T})^{-2}$.
Such a result clearly reflects the influence of the two new Kondo temperatures on the tunneling resistance. Also, for normal Kondo coupling $H_K^{(1)}$, it contributes the relaxation time $\tau_0$ with $\tau_0^{-1}\sim\frac{3\pi}{16\rho}[-1+2(1+\ln\frac{T_K^{(0)}}{T})^{-2}]$. Thus, the total relaxation time follows the result of $\tau_F^{-1}\sim \sum_{j=0}^2c_j(1+\ln\frac{T_K^{(j)}}{T})^{-2}$. This means that with the temperature decrease, three transition points will appear in the process of resistance change. Meanwhile, we should notice that in addition to $H_K^{(2)}$, $H_T$ contributes to the Andreev reflection between the Majorana doublet and metal. Therefore, the resistance increase will be suppressed to some extent with the decrease of temperature.

\par
\emph{Conclusion}.
---To conclude, we have investigated the Kondo effect in one hetero-structure where one KQD couple to a normal metal and a \emph{DIII}-class TS contributing Majorana doublets at its ends. As a result, it has been found that two Kondo correlations co-occur between the localized state in the KQD and two continuum states, i.e., the continuum state in the normal metal and the continuum Andreev reflection state between the normal metal and TS. This exactly leads to the complication of the Kondo effect, including the special symmetry of the Kondo Hamiltonian and the presence of three Kondo temperatures. Consequently, with the change of the KQD level or Majorana-KQD coupling, this structure can be characterized by different Kondo temperatures. In contrast, all the results will disappear when the TS degrades to $D$-class. Therefore, we ascertain that the change of Kondo physics exactly reflects the fundamental difference between the Majorana modes in \emph{D}-class and \emph{DIII}-class TSs.

\par

\emph{Acknowlegdments}. 
---The authors are grateful to Yi-Zhuang You for helpful discussions. This work was financially supported by the Fundamental
Research Funds for the Central Universities (Grant
No. N130505001), the Natural Science
Foundation of Liaoning province of China (Grant
No. 2013020030), and the Liaoning BaiQianWan Talents
Program (Grant No. 2012921078).

\clearpage

\bigskip

\end{document}